# Braid analysis of a bouncing ball

Nicholas B. Tufillaro
*Center for Nonlinear Studies and Theoretical Division (T-13), MS-B258*
*Los Alamos National Laboratory, Los Alamos, NM 87545 USA*
(5 September 1994)

I identify the template organizing the chaotic dynamics of a bouncing ball system. I further show how to estimate the topological parameter values of the system directly from a time series—a process I call "topological time series analysis." Two distinct methods to determine the topological parameters are illustrated and compared—the "pruning front" procedure and a "braid analysis." Both procedures lead to compatible results.

05.45.+b, 47.52.+j

## I. INTRODUCTION

Braids arise as periodic orbits in dynamical systems modeled by three-dimensional flows [1–4]. The existence of a single periodic orbit of a dynamical system can imply the coexistence of many other periodic orbits [5–8]. The most well-known example of this phenomenon occurs in the field of one-dimensional dynamics and is described by Sarkovskii's Theorem [9]. Less well-known is the fact that analogous results hold for two-dimensional systems [3]. In one-dimensional dynamics it is useful to study the period (or the permutation) of an orbit [10]. In two-dimensional systems it is useful to study the *braid type* of an orbit [2]. Given this specification, we can ask whether or not the existence of a given braid (periodic orbit) *forces* the existence of another; as in the one-dimensional case, algorithms have recently been developed for answering this question [11–13].

As originally observed by Auerbach et. al., unstable periodic orbits are available in abundance from a single chaotic time series using the method of close recurrence [8,14,15]. By a "braid analysis" I propose to analyze a chaotic time series by first extracting an (incomplete) spectrum of periodic orbits, and second ordering the extracted orbits according to their orbit forcing relationship. As shown in this paper, it is often possible to find a single periodic orbit, or a small collection of orbits, which forces many orbits in the observed spectrum. These orbits also force additional orbits of arbitrarily high period. This analysis is restricted to "low-dimensional" flows (roughly, flows which can be modeled by systems with one unstable Lyapunov exponent), however it has a strong predictive capability.

I would also like to point out that this analysis gives us an effective and mathematically well defined "pruning procedure" for chaotic two-dimensional diffeomorphisms [16]. Instead of asking for rules describing which orbits are missing (pruned), I instead look for those orbits which must be present. For low period orbits (say up to period 11) this procedure can predict all those orbits which must be present in the flow. This procedure will usually miss orbits of higher period, however from an experimental viewpoint the low period orbits are the most important and accessible. Orbits of low period often force an infinity of other orbits. This is illustrated in one-dimensional dynamics by the famous statement "period three implies chaos" [17]. An analogous statement in two-dimensional dynamics is that a non-well-ordered period three braid implies chaos [18].

This paper is organized as follows. Section II reviews the dynamics of the bouncing ball system. In section III I identify the template organizing the chaotic flow of the bouncing ball system. It is a horseshoe with a full twist. In section IV I show how braid analysis works by applying it to times series data generated from the bouncing ball system. The analysis builds directly on the original topological analysis of such data sets due to Mindlin et. al. [19]. This section also illustrates how *easily measured braid invariants of the periodic orbits lead to strong dynamical information about the flow*—without the need for a problematic and detailed symbolic description of the orbits in phase space. Section V illustrates how a pruning front can be estimated from a collection of low period orbits. The results are compatible with the braid analysis of the previous section. Section VI offers some concluding remarks.

In the example studied in this paper I do have good control of the symbolics. In principle, though, a braid analysis does not require good control of the symbolics (a good partition) and can thus overcome some of the current difficulties associated with finding good symbolic descriptions for (nonhyperbolic) strange attractors [20].

## II. BOUNCING BALL SYSTEM

Consider the motion of a ball bouncing on a periodically vibrating table. This bouncing ball system arises quite naturally as a model problem in several engineering applications: examples include the generation and control of noise in machinery such as jackhammers, the transportation and separation of granular solids such as rice, and the transportation of components in automatic assembly devices which commonly employ oscillating tracks. Several researchers have studied one-dimensional models of the bouncing ball system which include the coefficient of restitution ($0 \leq \alpha \leq 1$), and many have also noted the existence of a large class periodic, chaotic, and eventually periodic orbits known as "sticking solutions" [21]. More details can be found in Ref. [22]. All these models have been termed the "exact" one-dimensional model of the bouncing ball system. The phrase "one-dimensional" refers to the number of degrees of freedom the ball moves in and not to the dimension of the phase space model.

To fix a notation which allows an easier comparison with experiments, recall that the dynamics of the bouncing ball system can be found by solving the (implicit) nonlinear coupled algebraic equations known as the *phase map*,

$$A[\sin(\theta_k) + 1] + v_k \left[\frac{1}{\omega}(\theta_{k+1} - \theta_k)\right]$$
$$-\frac{1}{2}g\left[\frac{1}{\omega}(\theta_{k+1} - \theta_k)\right]^2 - A[\sin(\theta_{k+1}) + 1] = 0 \qquad (1)$$

and the *velocity map*,

$$(1 + \alpha)\omega A \cos(\theta_{k+1}) - \alpha \left\{v_k - g\left[\frac{1}{\omega}(\theta_{k+1} - \theta_k)\right]\right\} = v_{k+1} \qquad (2)$$

where $\theta_k = \omega t + \theta_0$ and $v_k$ are the phase and velocity of the k-th impact between the ball and oscillating table, $A$ and $\omega$ are the table's amplitude and angular frequency, $\alpha$ is the coefficient of restitution, and $g$ is the gravitational acceleration. The table's forcing period is denoted by $T = 2\pi/\omega$. The implicit phase map and explicit velocity map constitute the *exact* model of the bouncing ball system. Earlier experimental studies showed an excellent correspondence between the exact model and the dynamics of an experimental bouncing ball system, all the major bifurcations predicted by the exact model occurred within the experimental system [23]. Observations between the model and experiment agreed to within %2 with no fitted parameters. A public domain program, the *Bouncing Ball Simulation System*, is available to simulate the exact model [22]. I use this program to obtain the time series data analyzed here.

Experiments illustrating chaos in the bouncing ball system usually proceed along the following lines. The amplitude of the table driving the ball is slowly increased while monitoring the dynamics of the bouncing ball through an experimental impact map, which is similar to a next return map. In essence, an experimental bifurcation diagram is created. The coefficient of restitution can be changed from around 0.2 to 0.8 by using different materials for the ball (eg., wood, plastic, steel). Experimentally, it is observed that a chaotic invariant set is seen at the end of the period doubling cascade, but for a further increase in the driving amplitude, the strange attractor is destroyed by a crisis [24]. The dynamics of the ball after this crisis can result in motion which can quickly approach a periodic sticking solution (generally speaking, for smaller values of $\alpha$), or can exhibit long transients—sometimes called 'transient chaos' [25]— following the "shadow of the strange attractor" (generally speaking, for larger values of $\alpha$).

Direct simulation of the exact model exhibits a similar behavior. Figure 1 presents a bifurcation diagram showing a period doubling route to chaos for $\alpha = 0.5$. Note that this strange attractor is approached in the same way as it would be in an experiment, namely, by slowly scanning the amplitude until the end of the period doubling cascade is reached and a non-periodic orbit is observed. In simulations ($A = 0.012$) the strange attractor is found to be stable for over $10^6$ impacts. For $A > 0.0.118$ and $A < 0.019$ a first crisis occurs which expands the size of the strange attractor. Between $A = 0.0121$ and $A = 0.0122$ a second crisis occurs which destroys this strange attractor. For $A > 0.0122$ the orbit follows the shadow of the strange attractor for a number of impacts but eventually converges to a sticking solution (typically after $10^2$ to $10^3$ impacts). In both experiments and simulations, the pre-crisis (chaotic) dynamics and post-crisis (eventually periodic) dynamics are usually easy to distinguish because the range of impact phases explored by the ball suddenly widens after the crisis. In the simulation shown in Figure 1, the chaotic dynamics is confined to a phase between $-0.1 < \theta/2\pi < 0.3$ where as the (second) post-crisis dynamics explores almost the entire range of phases available. This feature provides a nice signature to distinguish the pre- and post-crisis dynamics in both experiments and simulations.

This general scenario of period-doubling, chaos, crisis, and sticking solutions (possibly with transient chaos) is not confined to a few selected parameter values but is generally observed for a wide range of $\alpha$. Both examples of crisis occurring in the bouncing ball system, and the existence of multiple coexisting attractors, can not be explained by a one dimensional unimodal theory, and provide the first indication of the need for a two dimensional theory.

### III. BOUNCING BALL TEMPLATE

The first step in analyzing the structure of the chaotic set in the bouncing ball system is the identification of a template which captures the organization of its periodic orbits [19]. The template is a nice (canonical) representation of the stretching, folding, and twisting of phase space resulting in a particular chaotic form. To visualize the template arising in the bouncing ball system I plot a chaotic orbit in the three dimensional space, $(\sin(\omega t), v(t), x(t))$, where the first coordinate is the table's (normalized) time dependent forcing amplitude, and the remaining coordinates are the ball's velocity and height.

Inspection of Fig. 2(a) reveals a band like structure with a half twist occurring where the ball reverses velocity when it hits the table, and an additional smaller fold on the outer edge of the band near the top of the figure. A schematic of the sheet like structure is presented in Fig. 2(b). A template is nothing more or less than this sheeted structure collapsed to a single sheet and moved by a sequence of isotopies to a standard form. This sheeted structure is perhaps easier to see in Fig. 3. Here the pre-image of the fold can be traced back to its impact point with the table. The impact phase of the fold point is in the vicinity of $\theta_k \approx 0$. The folding of the strange attractor in phase space occurs in the region where the table's impact velocity is maximal: roughly, orbits hitting at phases greater or lesser than this value get less of a kick from the table and hence do not travel as high.

Figure 4 shows how this sheeted structure can be put into a template of standard form. In Fig. 4(a) the evolution of a small section of an unstable manifold (represented by an arrow) is shown as it is carried by the template. To reach a canonical form I first pull the fold point all the way to the left (thus going from a pruned to unpruned system), and second identify and cut through the trajectory of the fold point. In the language of Cvitanović et. al. [5], this fold point is a primary tangency. As shown in Fig. 4(b) each branch of the template is now given a symbolic label. I also put the insertion layer of the template in standard form (back to front) and slide all the branch twists to the top of the diagram [4]. The template of the bouncing ball system in standard form is shown in Fig. 4(c). At this point I notice that by subtracting a full twist from the entire template I arrive at a horseshoe template in standard form (Fig. 4(d))—thus, the template in the bouncing ball system for the parameter range considered is the horseshoe with a global torsion of $-1$. In the next section I verify that the template identified is correct by comparing topological invariants calculated from the horseshoe template and those extracted directly from a chaotic time series.

## IV. BRAID ANALYSIS

A braid analysis of a low dimensional chaotic time series consists of four steps once an appropriate three-dimensional space is created [19]: (i) the periodic orbits are extracted by the method of close recurrence [22,26], (ii) the braid type of each periodic orbit is identified and the orbits are ordered by their two-dimensional forcing relationship [11,27] (iii) a subset of braids are selected which have maximal forcing and which force the orbits extracted in step (i), and (iv) if possible, an attempt is made to verify that some of the predicted orbits (not originally extracted in step (i)) are found in the system.

In practice, steps (i) and (ii) are greatly simplified if the template or knot-holder organizing the flow can be identified using the procedure described by Mindlin and co-workers [4,19,28,29]. Knowledge of the template helps in obtaining the symbolic names of the periodic orbits and in calculating the forcing relationship for the specific braids in that template. For instance, if the template is identified as a two-branch horseshoe knot holder (as is the example studied in this paper), then the theory of quasi-one-dimensional (qod) orbits of Hall [6,27] can be applied to simplify the analysis.

Although template identification is very valuable, it is not essential for a braid analysis. Nor is the symbolic identification of the extracted orbits. In the worst case a braid analysis does require that the the braid conjugacy class of each extracted periodic orbit is identified (see Elrifai and Morton [30], or Jaquemard [31] for algorithms), and that the minimal Markov model (a 'train track' in the language of Thurston) can be constructed for each braid (see Bestvina and Handel [11], Los [12], and Franks and Misiurewicz [13] for algorithms). Algorithms exist for both of these steps, although the most computationally efficient version of the braid conjugacy algorithm is probably not an effective solution for braids beyond period 8.

To illustrate braid analysis I simulated the bouncing ball system for $10^5$ impacts with system parameters $\alpha = 0.5$, $\omega = 2\pi 60$, and $A = 0.01215$. The resulting next impact map, $(\phi_k, \phi_{k+1})$, $\phi_k = \theta_k/2\pi$, is shown in Fig. 5. The many leaves of this return map once again indicates that the symbolic dynamics of this system should exhibit departures from that predicted by a one dimensional unimodal theory. The inset of the Fig. 5 shows an expand view of the region surrounding the maximum of the map. Three distinct leaves are visible in this region and this suggests that, to a first approximation, the symbolic dynamics of the system should be describable by a a three step pruning front.

To extract the (approximate) periodic orbits by the method of close recurrence I first convert the next impact map from the sequence of values $(\phi_k, \phi_{k+1})$ directly into a symbol sequence of $0's$ and $1's$. In this particular instance, I found that an adequate symbolic description (at least up to period 11 orbits, or approximately one part in $2^{11}$) is obtained by choosing the maximum value of the next impact map at the three leaves shown in Fig. 5. Orbits passing to the left of the maximum in the vicinity of a given layer are labeled zero, and those to the right are labeled one. Next I search this symbolic encoding for each and every periodic symbol string. Every time a periodic symbol string is found I calculate its (normalized) recurrence and then save the instance of the orbit with the best recurrence. The advantage of this procedure of orbit extraction is that it is exhaustive. I search for every possible orbit up to a given period. In these studies I searched for all orbits between periods 1 through 11. Some of the extracted periodic orbits are shown in Fig. 6.

The resulting spectrum of periodic orbits up to period 11 is shown in Table I. Simple topological invariants

(linking numbers and relative rotation rates) of the extracted orbits are calculated and compared with those of a horseshoe with a global torsion of -1. There are no discrepancies. This indicates that—at least to this level of resolution—the template is correctly identified and the symbolic partition is adequate. The orbits which are present in (the full shift) complete hyperbolic system, and not present in the Tables in Appendix A, are said to be *pruned*. Our goal is to predict as best as possible the pruned spectrum from the chaotic time series.

The symbolic label (up to braid type) can also be determined by considering simple and easily computable braid invariants. For instance, as pointed out by Hall [27], the exponent sum (simply the sum of braid crossings in the example) is a complete invariant for horseshoe braids up to period 8 (see Table II). Also, as inspection of the tables in Appendix A reveals, the exponent sum manages to distinguish most of the the pseudo-Anosov horseshoe braids of periods 9, 10, and 11 as well. Thus, I see that an easily determined quantity measured from a time series leads to the conclusion that the flow contains a chaotic invariant set—with out the calculation of more detailed quantities such as fractal dimensions or Lyapunov exponents.

The goal of a braid analysis is to find a small subset of orbits, called a "basis set" [7], which forces the the observed periodic orbit spectrum. One sensible way to proceed in identifying such a collection of orbits is to calculate the spectrum of orbits forced by a few high entropy orbits to see if they can capture most or all of the observed spectrum. If some orbits are left out then they are systematically added to the basis set until all the orbits in the observed spectrum are captured.

Using the tables in Appendix A I find that the highest entropy orbit in this particular data set is $s_9^2$ ($h_2 = 0.397$) which happens to be a quasi-one-dimensional orbit. Thus, using the results of Hall [6,27], I calculate the forced spectrum of this orbit by one-dimensional unimodal kneading theory [32]. I find that up to period eleven the $s_9^2$ braids forces $s_{11}^3$ ($s_{11}^4$), $s_7^1$, $s_{11}^1$, $s_9^1$, $s_{11}^1$, $s_{10}^3$, $s_8^2$, $s_{10}^2$, $s_6^1$, $s_{10}^1$, $s_8^1$, $s_4^1$, $s_2^1$, $s_1^1$. There are still many orbits unaccounted for in the observed spectrum. I thus examine the orbit with the next highest entropy. It is the $s_9^4$ ($h_2 = 0.377$) orbit. The $s_9^4$ braid is not QOD. Its forced spectrum can be calculated either by obtaining the train track for this braid (and the associated Markov model) [11], or by the method of pruning fronts briefly illustrated in section V. Using a train track calculation I find that the $s_9^4$ braid forces $s_{11}^5$ ($s_{11}^8$), $s_{10}^4$ ($s_{10}^5$), $s_5^1$, $s_{11}^4$ ($s_{11}^3$), $s_7^1$, $s_9^1$, $s_{11}^1$, $s_{10}^3$, $s_8^2$, $s_6^1$, $s_2^1$, $s_1^1$. Comparing the spectrum forced by the union of these two orbits with the observed spectrum (Table I), I find that only one orbit is unaccounted for, the finite order braid $s_8^3$, which is the maximal orbit in the observed data set in terms of one-dimensional unimodal theory.

Adding this orbit to the collection I determine that a basis set which accounts for the observed spectrum—up to braid type—is $\{s_8^3, s_9^4, s_9^2\}$

## V. PRUNING FRONTS FROM PERIODIC ORBITS

I now attempt to predict the (low period) forced orbits by using the extracted periodic orbits to systematically construct an approximation to the pruning front [5]. The braid analysis of the previous section only specifies the existence of orbits up to braid type. Thus, for instance, it might only predict the existence of one individual periodic orbit in a given saddle-node pair. The pruning front procedure is more specific, it actually forces individual periodic orbits as denoted by their complete symbolic label. Not unexpectedly, therefore, the "basis set" of periodic orbits needed to constructed an approximate pruning front may be larger than that found in a braid analysis.

As a first step in obtaining an approximate pruning front I plot the trajectory of a single chaotic orbit in the symbol plane [5]. The data is a symbolic symbol string constructed in the previous section of the form

$$s = \ldots s_{-3}s_{-2}s_{-1}s_0.s_1s_2s_3\ldots$$

where symbols to the left and right of $s_0$ are the past and future symbols respectively. The coordinates of the symbol plane for a horseshoe are calculated from the well ordered past ($c_i$) and future ($b_i$) symbols as follows:

$$x(s) = \sum_{i=1}^{D} \frac{b_i}{2^i}, \quad b_i = \sum_{j=1}^{i} s_j \mod 2$$

and

$$y(s) = \sum_{i=0}^{D-1} \frac{c_i}{2^i}, \quad c_i = \sum_{j=0}^{i+1} s_{-j} \mod 2.$$

If s is an infinite symbol string generated by a chaotic orbit, then $D$ is infinity in the above sums. However, since I am dealing with finite data sets, I approximate the symbol plane coordinates of a point by taking $D = 16$. In this way I can use a finite symbol string from a chaotic trajectory to generate a sequence of points on the symbol plane. The resulting plot for the data is shown in Fig. 7.

An expanded view of the primary pruning front region (center right of full diagram) is shown in the inset of Fig 7. The two dimensional nature of the data set is indicated the by the steps in the diagram. If the data set was one dimensional then a vertical pruning front with no steps would be seen. Such vertical pruning fronts are found, for instance, when the bouncing ball system is much more dissipative ($\alpha \approx 0.1$). As a rule of thumb, the depth of the steps increases as the dissipation decreases. In this example, the steps are easier to see in the iterates of the pruning front.

Now, to construct an approximate pruning front I plot all the periodic orbits (periods one through eleven) extracted in the previous section on the symbol plane and examine their location in the region of the pruning front. This is shown in Fig. 8 in the same region as that found in the inset of Fig. 7. The periodic orbits closest to the right and the center (i.e. closest to the pruning front suggested by Fig. 7) are selected as a basis set for constructing an approximate pruning front. Labels for these innermost periodic orbits are indicated in Fig. 8, and the last digit in the symbolic label (for the saddle-node partners) is determined by whether rightmost point of the orbit lands above (1) or below (0) a line through the center of the symbol plane. To construct the approximate pruning front I take the orbit in each saddle-node pair which is larger (rightmost) by unimodal ordering. Thus in this example I construct the pruning front from the set of periodic orbits (from bottom right to center): $s_8^3(0)$, $s_{10}^5(0)$, $s_{11}^8(0)$, $s_9^4(0)$, $s_9^2(1)$.

An approximate pruning front is then constructed from a continuous sequence of horizontal and vertical line segments connecting these periodic orbits:

$v_0[s_8^3] \rightarrow$
$(\overline{0.10110110}, \overline{01101101.10110110})$,
$h_1[s_8^3, s_{10}^5] \rightarrow$
$(\overline{01101101.10110110}, \overline{01101101.1011011110})$,
$v_1[s_8^3, s_{10}^5] \rightarrow$
$(\overline{01101101.1011011110}, \overline{0111101101.1011011110})$,
$h_2[s_{10}^5, s_{11}^8] \rightarrow$
$(\overline{0111101101.1011011110}, \overline{0111101101.10110111110})$,
$v_2[s_{10}^5, s_{11}^8] \rightarrow$
$(\overline{0111101101.10110111110}, \overline{01111101101.10110111110})$,
$h_3[s_{11}^8, s_9^4] \rightarrow$
$(\overline{01111101101.10110111110}, \overline{01111101101.101101110})$,
$v_3[s_{11}^8, s_9^4] \rightarrow$
$(\overline{01111101101.101101110}, \overline{011101101.101101110})$,
$h_4[s_9^4, s_9^2] \rightarrow$
$(\overline{011101101.101101110}, \overline{011101101.101111011})$,
$v_4[s_9^4, s_9^2] \rightarrow$
$(\overline{011101101.101111011}, \overline{110111101.101111011})$.

By construction, this pruning front generates the same periodic orbit spectrum—up to period eleven—as that recorded in Table 1.

Like the braid analysis, beyond period eleven this pruning front should begin to generated fewer orbits than are actually present. Both of the pruning methods illustrated here are systematic approximations in the sense that, given a periodic orbit spectrum up to period $P$, these methods generate an exact spectrum up to some period $Q$, beyond which both methods then provide lower bounds on the periodic orbit spectrum.

## VI. CONCLUSION

I have illustrated how to determine the topological form (the template) and how to estimate topological parameters directly from a chaotic time series generated by a dissipative bouncing ball system. Two distinct techniques where used to predict orbit forcing—a braid analysis, and the pruning front approach. Both techniques provide an effective procedure for calculating the orbit spectrum of low period orbits. Both procedures also provide information (a lower bound) for the spectrum of all periodic orbits.

Each procedure for estimating the periodic spectrum has distinct advantages and disadvantages. The braid analysis does not require a symbolic partition, and is thus useful in the cases (eg. low dissipation) where determining an exact, or approximate symbolic partition, is problematic. The braid analysis is also based on a rigorous mathematical foundation. The braid analysis, though, only provides information about orbit forcing up to braid type. The chief advantage of the pruning front approach is that it provides information about individual periodic orbits. Its chief disadvantage is that it, as yet, rests on a weaker mathematical foundation and requires the construction of a symbolic partition.

In retrospect, I find it remarkable that such a small subset of periodic orbits (which are rather easy to get from experiments) contain so much topological and dynamical information about a (low-dimensional) flow. A few low period orbits are sufficient to determine the template describing the stretching and folding of the strange set. The template provides an upper bound to the topological entropy and is, in a sense, a maximally (i.e., a full shift) hyperbolic set which can be formally associated to a (possibly nonhyperbolic) strange set. In this paper I show how periodic orbits (and their associated hyperbolic sets) can be used to obtain an approximation to a strange set which is probably not hyperbolic. Formally, I might say that the hyperbolic set associated with each pseudo-Anosov braid is embedded within the strange attractor I am trying to describe in the sense that the (possibly nonhyperbolic) strange set must contain at least all the orbits forced by the extracted pseudo-Anosov braid.

## ACKNOWLEDGMENTS

I thank Toby Hall for his explanations of QOD orbit theory, braids, horseshoes, train tracks, and for programs used in calculating the Hall invariants of horseshoe braids; Tad White for providing a copy of his Foldtool program for calculating train tracks from a free group; and Bob Gilmore whose questions began these inquiries.

I also thank André de Carvalho for extensive discussions on pruning fronts—André recently announced theorems which firm up the mathematical foundations for pruning fronts.

# APPENDIX A: INVARIANTS

Topological invariants of horseshoe braids from period one through eleven are shown.

---

FIG. 1. Bifurcation diagram for the bouncing ball system.

FIG. 2. (a) Three dimensional plot of a chaotic trajectory. (b) Schematic of the sheeted structure governing the evolution of the chaotic trajectory.

FIG. 3. Two dimensional plot of a chaotic trajectory in the bouncing ball system. Inset shows template governing the evolution of the orbits.

FIG. 4. Transformations taking the template found in the phase space to a template in standard form.

FIG. 5. Next phase map for the bouncing ball system. Inset shows expanded view of the region near the maximum.

FIG. 6. Periodic orbits extracted from a chaotic time series. The exponent sum identifies the orbit up to braid type. The linking number of the $l(0, 10)$ orbits is also indicated.

FIG. 7. Symbol plane diagram generated by a chaotic trajectory in the bouncing ball system. Inset suggests an outline for a pruning front.

FIG. 8. Extracted periodic orbits, up to period eleven, plotted on the symbol plane. Taking the innermost points leads to an approximation of the pruning front.

TABLE I. Spectrum of low period orbits extracted from a chaotic time series of the bouncing ball system (all orbits with $\epsilon < 0.01$) are shown). Extracted orbits and their (best) normalized recurrence are recorded. Note that in this particular example all saddle-node partners are detected in pairs.

| P | $c_P$ | $\epsilon(\phi_k, \phi_{k+P})$ | P | $c_P$ | $\epsilon(\phi_k, \phi_{k+P})$ |
|---|---|---|---|---|---|
| $s_1^1$ | 1 | 0.001315 | $s_{10}^1$ | 1011101010 | 0.000686 |
| $s_1^2$ | 10 | 0.000185 | $s_{10}^1$ | 1011101011 | 0.001030 |
| $s_4^1$ | 1011 | 0.000265 | $s_{10}^2$ | 1011111010 | 0.000771 |
| $s_5^1$ | 10110 | 0.000254 | $s_{10}^2$ | 1011111011 | 0.001599 |
| $s_5^1$ | 10111 | 0.000546 | $s_{10}^3$ | 1011111110 | 0.001274 |
| $s_6^1$ | 101110 | 0.000347 | $s_{10}^3$ | 1011111111 | 0.000510 |
| $s_6^1$ | 101111 | 0.000119 | $s_{10}^4$ | 1011010111 | 0.001372 |
| $s_7^1$ | 1011110 | 0.000262 | $s_{10}^5$ | 1011011110 | 0.004542 |
| $s_7^1$ | 1011111 | 0.000585 | $s_{10}^5$ | 1011011111 | 0.003181 |
| $s_8^1$ | 10111010 | 0.000069 | $s_{11}^1$ | 10111111110 | 0.002397 |
| $s_8^2$ | 10111110 | 0.000396 | $s_{11}^1$ | 10111111111 | 0.002669 |
| $s_8^2$ | 10111111 | 0.001261 | $s_{11}^2$ | 10111111010 | 0.005078 |
| $s_8^3$ | 10110110 | 0.000033 | $s_{11}^2$ | 10111111011 | 0.001549 |
| $s_8^3$ | 10110111 | 0.000097 | $s_{11}^3$ | 10111101010 | 0.000130 |
| $s_9^1$ | 101111110 | 0.000473 | $s_{11}^3$ | 10111101011 | 0.000372 |
| $s_9^1$ | 101111111 | 0.002288 | $s_{11}^4$ | 10111101110 | 0.000510 |
| $s_9^2$ | 101111010 | 0.000147 | $s_{11}^4$ | 10111101111 | 0.001238 |
| $s_9^2$ | 101111011 | 0.001211 | $s_{11}^5$ | 10110101110 | 0.002779 |
| $s_9^4$ | 101101110 | 0.000199 | $s_{11}^5$ | 10110101111 | 0.002854 |
| $s_9^4$ | 101101111 | 0.000086 | $s_{11}^8$ | 10110111110 | 0.002673 |
|  |  |  | $s_{11}^8$ | 10110111111 | 0.001730 |

TABLE II. Exponent sums for horseshoe braids up to period 8: standard horseshoe ($e_S(b)$) and horseshoe with a negative full twist ($e_S(b_{-1})$.) Orbits with the same exponent sum are braid conjugates. See [27] for the explicit conjugations.

| $c_P$ | $e_S(b)$ | $e_S(b_{-1})$ | $c_P$ | $e_S(b)$ | $e_S(b_{-1})$ |
|---|---|---|---|---|---|
| 0,1 | 0 | 0 | $101111_1^0$ | 18 | -24 |
| 10 | 1 | -1 | $101101_1^0$ | 16 | -26 |
| $10_1^0$ | 2 | -4 | $100101_1^0$ | 14 | -28 |
| 1011 | 5 | -7 | $100111_1^0$ | 14 | -28 |
| $100_1^0$ | 3 | -9 | $100110_1^0$ | 12 | -30 |
| $1011_1^0$ | 8 | -12 | $100010_1^0$ | 10 | -32 |
| $1001_1^0$ | 6 | -14 | $100011_1^0$ | 10 | -32 |
| $1000_1^0$ | 4 | -16 | $100001_1^0$ | 8 | -34 |
| $10111_1^0$ | 13 | -17 | $100000_1^0$ | 6 | -36 |
| 100101 | 9 | -21 | $1011111_1^0$ | 25 | -31 |
| $10011_1^0$ | 9 | -21 | 10111010 | 23 | -33 |
| $10001_1^0$ | 7 | -23 | $1011011_1^0$ | 21 | -35 |
| $10000_1^0$ | 5 | -25 | $1001011_1^0$ | 19 | -37 |
|  |  |  | $1001111_1^0$ | 19 | -37 |
|  |  |  | $1001010_1^0$ | 17 | -39 |
|  |  |  | $1001110_1^0$ | 17 | -39 |
|  |  |  | $1001101_1^0$ | 17 | -39 |
|  |  |  | $1000101_1^0$ | 15 | -41 |
|  |  |  | $1000111_1^0$ | 15 | -41 |
|  |  |  | 10001001 | 13 | -43 |
|  |  |  | $1000110_1^0$ | 13 | -43 |
|  |  |  | $1000010_1^0$ | 11 | -45 |
|  |  |  | $1000011_1^0$ | 11 | -45 |
|  |  |  | $1000001_1^0$ | 9 | -47 |
|  |  |  | $1000000_1^0$ | 7 | -49 |

TABLE III. Topological invariants for horseshoe braids up to period eleven (periods one through nine previously published by Hall [27]): name, code, permutation, Thurston type, rotation number, rotation interval, height, depth, exponent sum, and topological entropy of the braid calculated from a train track [11].

| P | $c_P$ | $\pi_P$ | Type | $\rho(P)$ | $\rho_i(P)$ | $q(P)$ | $r(P)$ | $e_S(P)$ | $h_2(P)$ |
|---|---|---|---|---|---|---|---|---|---|
| $s_1^1$ | 1 | (1) | fo | N/A | N/A | 1/2 | 1/2 | 0 | 0 |
| $s_2^1$ | 10 | (12) | fo | 1/2 | [1/2] | 1/2 | 1/2 | 1 | 0 |
| $s_3$ | $10_1^0$ | (123) | fo | 1/3 | [1/3] | 1/3 | 1/2 | 2 | 0 |
| $s_4^1$ | 1011 | (1324) | red | 1/2 | [1/2] | 1/2 | 1/2 | 5 | 0 |
| $s_4^4$ | $100_1^0$ | (1234) | fo | 1/4 | [1/4] | 1/4 | 1/2 | 3 | 0 |
| $s_5^1$ | $1011_1^0$ | (13425) | fo | 2/5 | [2/5] | 2/5 | 1/2 | 8 | 0 |
| $s_5^2$ | $1001_1^0$ | (12435) | pA | 2/5 | [1/3,1/2] | 1/3 | 1/3 | 6 | 0.544 |
| $s_5^3$ | $1000_1^0$ | (12345) | fo | 1/5 | [1/5] | 1/5 | 1/2 | 4 | 0 |
| $s_6^1$ | $10111_1^0$ | (143526) | red | 1/2 | [1/2] | 1/2 | 1/2 | 13 | 0 |
| $s_6^2$ | 100101 | (135246) | red | 1/3 | [1/3] | 1/3 | 1/2 | 9 | 0 |
| $s_6^3$ | $10011_1^0$ | (124536) | red | 1/3 | [1/3] | 1/3 | 1/2 | 9 | 0 |
| $s_6^4$ | $10000_1^0$ | (123546) | pA | 1/3 | [1/4,1/2] | 1/4 | 1/4 | 7 | 0.633 |
| $s_6^5$ | $10000_1^0$ | (123456) | fo | 1/6 | [1/6] | 1/6 | 1/2 | 5 | 0 |
| $s_7^1$ | $101111_1^0$ | (1453627) | fo | 3/7 | [3/7] | 3/7 | 1/2 | 18 | 0 |
| $s_7^2$ | $101101_1^0$ | (1462537) | pA | 3/7 | [2/5,1/2] | 2/5 | 2/5 | 16 | 0.442 |
| $s_7^3$ | $100101_1^0$ | (1362547) | pA | 3/7 | [1/3,1/2] | 1/3 | 1/2 | 14 | 0.477 |
| $s_7^4$ | $100111_1^0$ | (1254637) | pA | 3/7 | [1/3,1/2] | 1/3 | 1/2 | 14 | 0.477 |
| $s_7^5$ | $100110_1^0$ | (1356247) | fo | 2/7 | [2/7] | 2/7 | 1/2 | 12 | 0 |
| $s_7^6$ | $100010_1^0$ | (1246357) | pA | 2/7 | [1/4,1/3] | 1/4 | 1/2 | 10 | 0.382 |
| $s_7^7$ | $100011_1^0$ | (1235647) | pA | 2/7 | [1/4,1/3] | 1/4 | 1/2 | 10 | 0.382 |
| $s_7^8$ | $100001_1^0$ | (1234657) | pA | 2/7 | [1/5,1/2] | 1/5 | 1/5 | 8 | 0.666 |
| $s_7^9$ | $100000_1^0$ | (1234567) | fo | 1/7 | [1/7] | 1/7 | 1/2 | 6 | 0 |
| $s_8^1$ | 10111010 | (15472638) | red | 1/2 | [1/2] | 1/2 | 1/2 | 23 | 0 |
| $s_8^2$ | $1011111_1^0$ | (15463728) | red | 1/2 | [1/2] | 1/2 | 1/2 | 25 | 0 |
| $s_8^3$ | $1011011_1^1$ | (14725638) | fo | 3/8 | [3/8] | 3/8 | 1/2 | 21 | 0 |
| $s_8^4$ | $1001011_1^0$ | (13725648) | pA | 3/8 | [1/3,2/5] | 1/3 | 1/2 | 19 | 0.346 |
| $s_8^5$ | $1001010_1^0$ | (13647258) | pA | 3/8 | [1/3,1/2] | 1/3 | 1/3 | 17 | 0.498 |
| $s_8^6$ | $1001110_1^0$ | (13657248) | pA | 3/8 | [1/3,1/2] | 1/3 | 1/3 | 17 | 0.498 |
| $s_8^7$ | $1001111_1^1$ | (12564738) | pA | 3/8 | [1/3,2/5] | 1/3 | 1/2 | 19 | 0.346 |
| $s_8^8$ | $1001101_1^1$ | (12573648) | pA | 3/8 | [1/3,1/2] | 1/3 | 1/3 | 17 | 0.498 |
| $s_8^9$ | 10001001 | (13572468) | red | 1/4 | [1/4] | 1/4 | 1/2 | 13 | 0 |
| $s_8^{10}$ | $1000101_1^0$ | (12473658) | pA | 3/8 | [1/4,1/2] | 1/4 | 1/2 | 15 | 0.569 |
| $s_8^{11}$ | $1000111_1^0$ | (12365746) | pA | 3/8 | [1/4,1/2] | 1/4 | 1/2 | 15 | 0.569 |
| $s_8^{11}$ | $1000110_1^0$ | (12467358) | red | 1/4 | [1/4] | 1/4 | 1/2 | 13 | 0 |
| $s_8^{12}$ | $1000010_1^0$ | (12357468) | pA | 1/4 | [1/5,1/3] | 1/5 | 1/2 | 11 | 0.459 |
| $s_8^{13}$ | $1000011_1^0$ | (12346758) | pA | 1/4 | [1/5,1/3] | 1/5 | 1/2 | 11 | 0.459 |
| $s_8^{14}$ | $1000001_1^0$ | (12345768) | pA | 1/4 | [1/6,1/2] | 1/6 | 1/6 | 9 | 0.680 |
| $s_8^{15}$ | $1000000_1^0$ | (12345678) | fo | 1/8 | [1/8] | 1/8 | 1/2 | 7 | 0 |

| P | $c_P$ | $\pi_P$ | Type | $\rho(P)$ | $\rho_i(P)$ | $q(P)$ | $r(P)$ | $e_S(P)$ | $h_2(P)$ |
|---|---|---|---|---|---|---|---|---|---|
| $s_9^1$ | $101111111_1^0$ | (156473829) | fo | 4/9 | [4/9] | 4/9 | 1/2 | 32 | 0 |
| $s_9^2$ | $101111101_1^0$ | (156482739) | pA | 4/9 | [3/7,1/2] | 3/7 | 3/7 | 30 | 0.397 |
| $s_9^3$ | $101110101_1^0$ | (157382649) | pA | 4/9 | [2/5,1/2] | 2/5 | 1/2 | 28 | 0.377 |
| $s_9^4$ | $101110111_1^0$ | (148265739) | pA | 4/9 | [2/5,1/2] | 2/5 | 1/2 | 28 | 0.377 |
| $s_9^5$ | $100010110_1^0$ | (147368259) | red | 1/3 | [1/3] | 1/3 | 1/2 | 22 | 0 |
| $s_9^6$ | $100010111_1^0$ | (138265749) | pA | 4/9 | [1/3,1/2] | 1/3 | 1/2 | 26 | 0.447 |
| $s_9^7$ | $100010101_1^0$ | (137482659) | pA | 4/9 | [1/3,1/2] | 1/3 | 1/2 | 24 | 0.507 |
| $s_9^8$ | $100011101_1^0$ | (126583749) | pA | 4/9 | [1/3,1/2] | 1/3 | 1/2 | 24 | 0.507 |
| $s_9^9$ | $100011111_1^0$ | (126574839) | pA | 4/9 | [1/3,1/2] | 1/3 | 1/2 | 26 | 0.447 |
| $s_9^{10}$ | $100111110_1^0$ | (136758249) | red | 1/3 | [1/3] | 1/3 | 1/2 | 22 | 0 |
| $s_9^{11}$ | $100110010_1^0$ | (136748259) | red | 1/3 | [1/3] | 1/3 | 1/2 | 22 | 0 |
| $s_9^{12}$ | $100110011_1^0$ | (125836749) | red | 1/3 | [1/3] | 1/3 | 1/2 | 22 | 0 |
| $s_9^{13}$ | $100110001_1^0$ | (136824759) | pA | 1/3 | [2/7,1/2] | 2/7 | 2/7 | 20 | 0.605 |
| $s_9^{14}$ | $100001001_1^0$ | (135824769) | pA | 1/3 | [1/4,1/2] | 1/4 | 1/3 | 18 | 0.537 |
| $s_9^{15}$ | $100001011_1^0$ | (124836759) | pA | 1/3 | [1/4,2/5] | 1/4 | 1/2 | 20 | 0.492 |
| $s_9^{16}$ | $100001010_1^0$ | (124758369) | pA | 1/3 | [1/4,1/2] | 1/4 | 1/3 | 18 | 0.537 |
| $s_9^{17}$ | $100001110_1^0$ | (124768359) | pA | 1/3 | [1/4,1/2] | 1/4 | 1/3 | 18 | 0.537 |
| $s_9^{18}$ | $100001111_1^0$ | (123675849) | pA | 1/3 | [1/4,2/5] | 1/4 | 1/2 | 20 | 0.492 |
| $s_9^{19}$ | $100001101_1^0$ | (123684759) | pA | 1/3 | [1/4,1/2] | 1/4 | 1/3 | 18 | 0.537 |
| $s_9^{20}$ | $100001100_1^0$ | (135782469) | fo | 2/9 | [2/9] | 2/9 | 1/2 | 16 | 0 |
| $s_9^{21}$ | $100000100_1^0$ | (124683579) | pA | 2/9 | [1/5,1/4] | 1/5 | 1/2 | 14 | 0.295 |
| $s_9^{22}$ | $100000101_1^0$ | (123584769) | pA | 1/3 | [1/5,1/2] | 1/5 | 1/2 | 16 | 0.605 |
| $s_9^{23}$ | $100000111_1^0$ | (123476859) | pA | 1/3 | [1/5,1/2] | 1/5 | 1/2 | 16 | 0.605 |
| $s_9^{24}$ | $100000110_1^0$ | (123578469) | pA | 2/9 | [1/5,1/4] | 1/5 | 1/2 | 14 | 0.295 |
| $s_9^{25}$ | $100000010_1^0$ | (123468579) | pA | 2/9 | [1/6,1/3] | 1/6 | 1/2 | 12 | 0.492 |
| $s_9^{26}$ | $100000011_1^0$ | (123457869) | pA | 2/9 | [1/6,1/3] | 1/6 | 1/2 | 12 | 0.492 |
| $s_9^{27}$ | $100000001_1^0$ | (123456879) | pA | 2/9 | [1/7,1/2] | 1/7 | 1/7 | 10 | 0.687 |
| $s_9^{28}$ | $100000000_1^0$ | (123456789) | fo | 1/9 | [1/9] | 1/9 | 1/2 | 8 | 0 |

| P | $c_P$ | $\pi_P$ | Type | $\rho(P)$ | $\rho_i(P)$ | $q(P)$ | $r(P)$ | $e_S(P)$ | $h_2(P)$ |
|---|---|---|---|---|---|---|---|---|---|
| $s_{10}^{1}$ | $1011101011^0$ | (16583927410) | red | 1/2 | [1/2] | 1/2 | 1/2 | 37 | 0 |
| $s_{10}^{2}$ | $1011111011^0$ | (16574928310) | red | 1/2 | [1/2] | 1/2 | 1/2 | 39 | 0.272 |
| $s_{10}^{3}$ | $1011111111^0$ | (16574839210) | red | 1/2 | [1/2] | 1/2 | 1/2 | 41 | 0 |
| $s_{10}^{4}$ | $1011010111^0$ | (15839267410) | red | 2/5 | [2/5] | 2/5 | 1/2 | 35 | 0 |
| $s_{10}^{5}$ | $1011011111^0$ | (14926758310) | red | 2/5 | [2/5] | 2/5 | 1/2 | 35 | 0 |
| $s_{10}^{6}$ | $1011011011^0$ | (15926837410) | pA | 2/5 | [3/8,1/2] | 3/8 | 3/8 | 33 | 0.473 |
| $s_{10}^{7}$ | $1001011011^0$ | (14926837510) | pA | 2/5 | [1/3,1/2] | 1/3 | 2/5 | 31 | 0.447 |
| $s_{10}^{8}$ | $1001011111^0$ | (13926758410) | pA | 2/5 | [1/3,3/7] | 1/3 | 1/2 | 33 | 0.394 |
| $s_{10}^{9}$ | $1001011101^0$ | (14837692510) | pA | 2/5 | [1/3,1/2] | 1/3 | 1/2 | 29 | 0.438 |
| $s_{10}^{10}$ | $1001010101^0$ | (14837592610) | pA | 2/5 | [1/3,1/2] | 1/3 | 1/2 | 29 | 0.438 |
| $s_{10}^{11}$ | $1001010111^0$ | (13849267510) | pA | 2/5 | [1/3,1/2] | 1/3 | 2/5 | 31 | 0.447 |
| $s_{10}^{12}$ | $1001110010$ | (13769248510) | red | 2/5 | [1/3,1/2] | 1/3 | 1/3 | 27 | 0.544 |
| $s_{10}^{13}$ | $1001110111^0$ | (13759268410) | pA | 2/5 | [1/3,1/2] | 1/3 | 2/5 | 31 | 0.447 |
| $s_{10}^{14}$ | $1001110101^0$ | (13768492510) | pA | 2/5 | [1/3, 1/2] | 1/3 | 1/2 | 29 | 0.438 |
| $s_{10}^{15}$ | $1001111101^0$ | (13768592410) | pA | 2/5 | [1/3,1/2] | 1/3 | 1/2 | 29 | 0.438 |
| $s_{10}^{16}$ | $1001111111^0$ | (12675849310) | pA | 2/5 | [1/3,3/7] | 1/3 | 1/2 | 33 | 0.394 |
| $s_{10}^{17}$ | $1001111101^0$ | (12675938410) | pA | 2/5 | [1/3,1/2] | 1/3 | 2/5 | 31 | 0.447 |
| $s_{10}^{18}$ | $1001101011^0$ | (12684937510) | pA | 2/5 | [1/3,1/2] | 1/3 | 1/2 | 29 | 0.438 |
| $s_{10}^{19}$ | $1001101111^0$ | (12593768410) | pA | 2/5 | [1/3,1/2] | 1/3 | 1/2 | 29 | 0.438 |
| $s_{10}^{20}$ | $1001101100^0$ | (14783692510) | fo | 3/10 | [3/10] | 3/10 | 1/2 | 27 | 0 |
| $s_{10}^{21}$ | $1001100010^0$ | (14792583610) | pA | 3/10 | [2/7,1/3] | 2/7 | 1/2 | 25 | 0.302 |
| $s_{10}^{22}$ | $1001100011^0$ | (13692478510) | pA | 3/10 | [2/7,1/3] | 2/7 | 1/2 | 25 | 0.302 |
| $s_{10}^{23}$ | $1000100011^0$ | (13592478610) | pA | 3/10 | [1/4,1/3] | 1/4 | 1/2 | 23 | 0.337 |
| $s_{10}^{24}$ | $1000100010^0$ | (13692584710) | pA | 3/10 | [1/4,1/3] | 1/4 | 1/2 | 23 | 0.337 |
| $s_{10}^{25}$ | $1000101100^0$ | (12584793610) | pA | 3/10 | [1/4,1/3] | 1/4 | 1/2 | 23 | 0.337 |
| $s_{10}^{26}$ | $1000101111^0$ | (12493768510) | pA | 2/5 | [1/4,1/2] | 1/4 | 1/2 | 27 | 0.544 |
| $s_{10}^{27}$ | $1000101011^0$ | (12485937610) | pA | 2/5 | [1/4,1/2] | 1/4 | 1/2 | 25 | 0.593 |
| $s_{10}^{28}$ | $1000101000^0$ | (13586924710) | pA | 3/10 | [1/4,1/2] | 1/4 | 1/4 | 21 | 0.612 |
| $s_{10}^{29}$ | $1000111000^0$ | (13587924610) | pA | 3/10 | [1/4,1/2] | 1/4 | 1/4 | 21 | 0.612 |
| $s_{10}^{30}$ | $1000111011^0$ | (12376948510) | pA | 2/5 | [1/4,1/2] | 1/4 | 1/2 | 25 | 0.593 |
| $s_{10}^{31}$ | $1000111111^0$ | (12376859410) | pA | 2/5 | [1/4,1/2] | 1/4 | 1/2 | 27 | 0.544 |
| $s_{10}^{32}$ | $1000111110^0$ | (12478693510) | pA | 3/10 | [1/4,1/3] | 1/4 | 1/2 | 23 | 0.337 |
| $s_{10}^{33}$ | $1000110010^0$ | (12478593610) | pA | 3/10 | [1/4,1/3] | 1/4 | 1/2 | 23 | 0.337 |
| $s_{10}^{34}$ | $1000110111^0$ | (12369478510) | pA | 3/10 | [1/4,1/3] | 1/4 | 1/2 | 23 | 0.337 |
| $s_{10}^{35}$ | $1000110001^0$ | (12479358610) | pA | 3/10 | [1/4,1/2] | 1/4 | 1/4 | 21 | 0.612 |
| $s_{10}^{36}$ | $1000010001$ | (13579246810) | red | 1/5 | [1/5] | 1/5 | 1/2 | 17 | 0 |
| $s_{10}^{37}$ | $1000010011^0$ | (12469358710) | pA | 3/10 | [1/5,1/2] | 1/5 | 1/3 | 19 | 0.559 |
| $s_{10}^{38}$ | $1000010111^0$ | (12359478610) | pA | 3/10 | [1/5,2/5] | 1/5 | 1/2 | 21 | 0.544 |
| $s_{10}^{39}$ | $1000010100^0$ | (12358694710) | pA | 3/10 | [1/5,1/2] | 1/5 | 1/3 | 19 | 0.559 |
| $s_{10}^{40}$ | $1000011100^0$ | (12358794610) | pA | 3/10 | [1/5,1/2] | 1/5 | 1/3 | 19 | 0.559 |
| $s_{10}^{41}$ | $1000011111^0$ | (12347869510) | pA | 3/10 | [1/5,2/5] | 1/5 | 1/2 | 21 | 0.544 |
| $s_{10}^{42}$ | $1000011101^0$ | (12347958610) | pA | 3/10 | [1/5,1/2] | 1/5 | 1/3 | 19 | 0.559 |
| $s_{10}^{43}$ | $1000011001^0$ | (12468935710) | red | 1/5 | [1/5] | 1/5 | 1/2 | 17 | 0 |
| $s_{10}^{44}$ | $1000001001^0$ | (12357946810) | pA | 1/5 | [1/6,1/4] | 1/6 | 1/2 | 15 | 0.362 |
| $s_{10}^{45}$ | $1000001011^0$ | (12346958710) | pA | 3/10 | [1/6,1/2] | 1/6 | 1/2 | 17 | 0.621 |
| $s_{10}^{46}$ | $1000001111^0$ | (12345879610) | pA | 3/10 | [1/6,1/2] | 1/6 | 1/2 | 17 | 0.621 |
| $s_{10}^{47}$ | $1000001101^0$ | (12346895710) | pA | 1/5 | [1/6,1/4] | 1/6 | 1/2 | 15 | 0.362 |
| $s_{10}^{48}$ | $1000000101^0$ | (12345796810) | pA | 1/5 | [1/7,1/3] | 1/7 | 1/2 | 13 | 0.508 |
| $s_{10}^{49}$ | $1000000111^0$ | (12345689710) | pA | 1/5 | [1/7,1/3] | 1/7 | 1/2 | 13 | 0.508 |
| $s_{10}^{50}$ | $1000000011^0$ | (12345679810) | pA | 1/5 | [1/8,1/2] | 1/8 | 1/8 | 11 | 0.690 |
| $s_{10}^{51}$ | $1000000000^0$ | (12345678910) | fo | 1/10 | [1/10] | 1/10 | 1/2 | 9 | 0 |

| P | $c_P$ | $\pi_P$ | Type | $\rho(P)$ | $\rho_i(P)$ | $q(P)$ | $r(P)$ | $e_S(P)$ | $h_2(P)$ |
|---|---|---|---|---|---|---|---|---|---|
| $s_{11}^{1}$ | $10111111111_1^0$ | (1675849310211) | fo | 5/11 | [5/11] | 5/11 | 1/2 | 50 | 0 |
| $s_{11}^{2}$ | $10111111101_1^0$ | (1675841029311) | pA | 5/11 | [4/9,1/2] | 4/9 | 4/9 | 48 | 0.374 |
| $s_{11}^{3}$ | $10111110101_1^0$ | (1675931028411) | pA | 5/11 | [3/7,1/2] | 3/7 | 1/2 | 46 | 0.331 |
| $s_{11}^{4}$ | $10111110111_1^0$ | (1684102759311) | pA | 5/11 | [3/7,1/2] | 3/7 | 1/2 | 46 | 0.331 |
| $s_{11}^{5}$ | $10110010111_1^0$ | (1593102768411) | pA | 5/11 | [2/5,1/2] | 2/5 | 1/2 | 44 | 0.344 |
| $s_{11}^{6}$ | $10110010101_1^0$ | (1693841027511) | pA | 5/11 | [2/5,1/2] | 2/5 | 1/2 | 42 | 0.412 |
| $s_{11}^{7}$ | $10110011101_1^0$ | (1510276938211) | pA | 5/11 | [2/5,1/2] | 2/5 | 1/2 | 42 | 0.412 |
| $s_{11}^{8}$ | $10110011111_1^0$ | (1410276859311) | pA | 5/11 | [2/5,1/2] | 2/5 | 1/2 | 44 | 0.344 |
| $s_{11}^{9}$ | $10110110011_1^0$ | (1510269378411) | fo | 4/11 | [4/11] | 4/11 | 1/2 | 40 | 0 |
| $s_{11}^{10}$ | $10010110011_1^0$ | (1410269378311) | pA | 4/11 | [1/3,3/8] | 1/3 | 1/2 | 38 | 0.288 |
| $s_{11}^{11}$ | $10010110101_1^0$ | (1493785102611) | pA | 4/11 | [1/3,2/5] | 1/3 | 1/2 | 36 | 0.302 |
| $s_{11}^{12}$ | $10010111101_1^0$ | (1493786102511) | pA | 4/11 | [1/3,2/5] | 1/3 | 1/2 | 36 | 0.302 |
| $s_{11}^{13}$ | $10010111111_1^0$ | (1310276859411) | pA | 5/11 | [1/3,1/2] | 1/3 | 1/2 | 42 | 0.432 |
| $s_{11}^{14}$ | $10010111101_1^0$ | (1410276938511) | pA | 5/11 | [1/3,1/2] | 1/3 | 1/2 | 40 | 0.466 |
| $s_{11}^{15}$ | $10010101011_1^0$ | (1493851027611) | pA | 5/11 | [1/3,1/2] | 1/3 | 1/2 | 38 | 0.497 |
| $s_{11}^{16}$ | $10010101111_1^0$ | (1394102768511) | pA | 5/11 | [1/3,1/2] | 1/3 | 1/3 | 40 | 0.466 |
| $s_{11}^{17}$ | $10010101100_1^0$ | (1485937102611) | pA | 4/11 | [1/3,1/2] | 1/3 | 1/3 | 34 | 0.486 |
| $s_{11}^{18}$ | $10010100010_1^0$ | (1486102593711) | pA | 4/11 | [1/3,1/2] | 1/3 | 1/3 | 32 | 0.510 |
| $s_{11}^{19}$ | $10011100010_1^0$ | (1487102593611) | pA | 4/11 | [1/3,1/2] | 1/3 | 1/3 | 32 | 0.510 |
| $s_{11}^{20}$ | $10011101100_1^0$ | (1487936102511) | pA | 4/11 | [1/3,1/2] | 1/3 | 1/3 | 34 | 0.486 |
| $s_{11}^{21}$ | $10011101111_1^0$ | (1385102769411) | pA | 5/11 | [1/3,1/2] | 1/3 | 1/2 | 40 | 0.466 |
| $s_{11}^{22}$ | $10011101011_1^0$ | (1276941038511) | pA | 5/11 | [1/3,1/2] | 1/3 | 1/2 | 38 | 0.497 |
| $s_{11}^{23}$ | $10011111011_1^0$ | (1276851039411) | pA | 5/11 | [1/3,1/2] | 1/3 | 1/2 | 40 | 0.466 |
| $s_{11}^{24}$ | $10011111111_1^0$ | (1276859410311) | pA | 5/11 | [1/3,1/2] | 1/3 | 1/2 | 42 | 0.432 |
| $s_{11}^{25}$ | $10011111110_1^0$ | (1378695102411) | pA | 4/11 | [1/3,2/5] | 1/3 | 1/2 | 36 | 0.302 |
| $s_{11}^{26}$ | $10011111010_1^0$ | (1378694102511) | pA | 4/11 | [1/3,2/5] | 1/3 | 1/2 | 36 | 0.302 |
| $s_{11}^{27}$ | $10011111011_1^0$ | (1378510269411) | pA | 4/11 | [1/3,3/8] | 1/3 | 1/2 | 38 | 0.288 |
| $s_{11}^{28}$ | $10011111001_1^0$ | (1378610249511) | pA | 4/11 | [1/3,1/2] | 1/3 | 1/3 | 34 | 0.486 |
| $s_{11}^{29}$ | $10011101001_1^0$ | (1379510248611) | pA | 4/11 | [1/3,1/2] | 1/3 | 1/3 | 32 | 0.510 |
| $s_{11}^{30}$ | $10011101011_1^0$ | (1269410378511) | pA | 4/11 | [1/3,2/5] | 1/3 | 1/2 | 36 | 0.302 |
| $s_{11}^{31}$ | $10011101010_1^0$ | (1379485102611) | pA | 4/11 | [1/3,1/2] | 1/3 | 1/3 | 34 | 0.486 |
| $s_{11}^{32}$ | $10011101110_1^0$ | (1379486102511) | pA | 4/11 | [1/3,1/2] | 1/3 | 1/3 | 34 | 0.486 |
| $s_{11}^{33}$ | $10011101111_1^0$ | (1251037869411) | pA | 4/11 | [1/3,2/5] | 1/3 | 1/2 | 36 | 0.302 |
| $s_{11}^{34}$ | $10011101101_1^0$ | (1261037948511) | pA | 4/11 | [1/3,1/2] | 1/3 | 1/3 | 34 | 0.486 |
| $s_{11}^{35}$ | $10011100101_1^0$ | (1371025948611) | pA | 4/11 | [2/7,1/2] | 2/7 | 1/2 | 32 | 0.538 |
| $s_{11}^{36}$ | $10011100111_1^0$ | (1361024879511) | pA | 4/11 | [2/7,1/2] | 2/7 | 1/2 | 32 | 0.538 |
| $s_{11}^{37}$ | $10011100110_1^0$ | (1471025893611) | fo | 3/11 | [3/11] | 3/11 | 1/2 | 30 | 0 |
| $s_{11}^{38}$ | $10000100110_1^0$ | (1361025894711) | pA | 3/11 | [1/4,2/7] | 1/4 | 1/2 | 28 | 0.254 |
| $s_{11}^{39}$ | $10000100111_1^0$ | (1351024879611) | pA | 4/11 | [1/4,1/2] | 1/4 | 1/2 | 30 | 0.486 |
| $s_{11}^{40}$ | $10000100101_1^0$ | (1361025948711) | pA | 4/11 | [1/4,1/2] | 1/4 | 1/2 | 30 | 0.486 |
| $s_{11}^{41}$ | $10000100100_1^0$ | (1369471025811) | pA | 3/11 | [1/4,1/3] | 1/4 | 1/2 | 26 | 0.348 |
| $s_{11}^{42}$ | $10000101100_1^0$ | (1369581024711) | pA | 3/11 | [1/4,1/3] | 1/4 | 1/2 | 26 | 0.348 |
| $s_{11}^{43}$ | $10000101101_1^0$ | (1251037948611) | pA | 4/11 | [1/4,1/2] | 1/4 | 2/5 | 32 | 0.513 |
| $s_{11}^{44}$ | $10000101111_1^0$ | (1241037869511) | pA | 4/11 | [1/4,3/7] | 1/4 | 1/2 | 34 | 0.517 |
| $s_{11}^{45}$ | $10000101110_1^0$ | (1259487103611) | pA | 4/11 | [1/4,1/2] | 1/4 | 1/2 | 30 | 0.486 |
| $s_{11}^{46}$ | $10000101010_1^0$ | (1259486103711) | pA | 4/11 | [1/4,1/2] | 1/4 | 1/2 | 30 | 0.486 |
| $s_{11}^{47}$ | $10000101011_1^0$ | (1249510378611) | pA | 4/11 | [1/4,1/2] | 1/4 | 2/5 | 32 | 0.513 |
| $s_{11}^{48}$ | $10000101001_1^0$ | (1359610248711) | pA | 4/11 | [1/4,1/2] | 1/4 | 1/3 | 28 | 0.629 |
| $s_{11}^{49}$ | $10000111001_1^0$ | (1248710359611) | pA | 4/11 | [1/4,1/2] | 1/4 | 1/3 | 28 | 0.629 |

| P | $c_P$ | $\pi_P$ | Type | $\rho(P)$ | $\rho_i(P)$ | $q(P)$ | $r(P)$ | $e_S(P)$ | $h_2(P)$ |
|---|---|---|---|---|---|---|---|---|---|
| $s_{11}^{50}$ | $1000111011_1^0$ | (1248610379511) | pA | 4/11 | [1/4,1/2] | 1/4 | 2/5 | 32 | 0.513 |
| $s_{11}^{51}$ | $1000111010_1^0$ | (1248795103611) | pA | 4/11 | [1/4,1/2] | 1/4 | 1/2 | 30 | 0.486 |
| $s_{11}^{52}$ | $1000111110_1^0$ | (1248796103511) | pA | 4/11 | [1/4,1/2] | 1/4 | 1/2 | 30 | 0.486 |
| $s_{11}^{53}$ | $1000111111_1^0$ | (1237869510411) | pA | 4/11 | [1/4,3/7] | 1/4 | 1/2 | 34 | 0.517 |
| $s_{11}^{54}$ | $1000111101_1^0$ | (1237861049511) | pA | 4/11 | [1/4,1/2] | 1/4 | 2/5 | 32 | 0.513 |
| $s_{11}^{55}$ | $1000111100_1^0$ | (1358971024611) | pA | 3/11 | [1/4,1/3] | 1/4 | 1/2 | 26 | 0.348 |
| $s_{11}^{56}$ | $1000110100_1^0$ | (1358961024711) | pA | 3/11 | [1/4,1/3] | 1/4 | 1/2 | 26 | 0.348 |
| $s_{11}^{57}$ | $1000110101_1^0$ | (1237951048611) | pA | 4/11 | [1/4,1/2] | 1/4 | 1/2 | 30 | 0.486 |
| $s_{11}^{58}$ | $1000110111_1^0$ | (1236104879511) | pA | 4/11 | [1/4,1/2] | 1/4 | 1/2 | 30 | 0.486 |
| $s_{11}^{59}$ | $1000110110_1^0$ | (1258947103611) | pA | 3/11 | [1/4,2/7] | 1/4 | 1/2 | 28 | 0.254 |
| $s_{11}^{60}$ | $1000110010_1^0$ | (1258103694711) | pA | 3/11 | [1/4,1/3] | 1/4 | 1/2 | 26 | 0.348 |
| $s_{11}^{61}$ | $1000110011_1^0$ | (1247103589611) | pA | 3/11 | [1/4,1/3] | 1/4 | 1/2 | 26 | 0.348 |
| $s_{11}^{62}$ | $1000110001_1^0$ | (1358102469711) | pA | 3/11 | [2/9,1/2] | 2/9 | 2/9 | 24 | 0.655 |
| $s_{11}^{63}$ | $1000010001_1^0$ | (1357102469811) | pA | 3/11 | [1/5,1/2] | 1/5 | 1/4 | 22 | 0.616 |
| $s_{11}^{64}$ | $1000010011_1^0$ | (1246103589711) | pA | 3/11 | [1/5,1/3] | 1/5 | 1/2 | 24 | 0.416 |
| $s_{11}^{65}$ | $1000010010_1^0$ | (1247103695811) | pA | 3/11 | [1/5,1/3] | 1/5 | 1/2 | 24 | 0.416 |
| $s_{11}^{66}$ | $1000010110_1^0$ | (1236958104711) | pA | 3/11 | [1/5,1/3] | 1/5 | 1/2 | 24 | 0.416 |
| $s_{11}^{67}$ | $1000010111_1^0$ | (1235104879611) | pA | 4/11 | [1/5,1/2] | 1/5 | 1/2 | 28 | 0.583 |
| $s_{11}^{68}$ | $1000010101_1^0$ | (1235961048711) | pA | 4/11 | [1/5,1/2] | 1/5 | 1/2 | 26 | 0.626 |
| $s_{11}^{69}$ | $1000010100_1^0$ | (1246971035811) | pA | 3/11 | [1/5,1/2] | 1/5 | 1/4 | 22 | 0.616 |
| $s_{11}^{70}$ | $1000011100_1^0$ | (1246981035711) | pA | 3/11 | [1/5,1/2] | 1/5 | 1/4 | 22 | 0.616 |
| $s_{11}^{71}$ | $1000011101_1^0$ | (1234871059611) | pA | 4/11 | [1/5,1/2] | 1/5 | 1/2 | 26 | 0.626 |
| $s_{11}^{72}$ | $1000011111_1^0$ | (1234879610511) | pA | 4/11 | [1/5,1/2] | 1/5 | 1/2 | 28 | 0.583 |
| $s_{11}^{73}$ | $1000011110_1^0$ | (1235897104611) | pA | 3/11 | [1/5,1/3] | 1/5 | 1/2 | 24 | 0.416 |
| $s_{11}^{74}$ | $1000011010_1^0$ | (1235896104711) | pA | 3/11 | [1/5,1/3] | 1/5 | 1/2 | 24 | 0.416 |
| $s_{11}^{75}$ | $1000011011_1^0$ | (1234710589611) | pA | 3/11 | [1/5,1/3] | 1/5 | 1/2 | 24 | 0.416 |
| $s_{11}^{76}$ | $1000011001_1^0$ | (1235810469711) | pA | 3/11 | [1/5,1/2] | 1/5 | 1/4 | 22 | 0.616 |
| $s_{11}^{77}$ | $1000011000_1^0$ | (1357910246811) | fo | 2/11 | [2/11] | 2/11 | 1/2 | 20 | 0 |
| $s_{11}^{78}$ | $1000001000_1^0$ | (1246810357911) | pA | 2/11 | [1/6,1/5] | 1/6 | 1/2 | 18 | 0.241 |
| $s_{11}^{79}$ | $1000001001_1^0$ | (1235710469811) | pA | 3/11 | [1/6,1/2] | 1/6 | 1/3 | 20 | 0.571 |
| $s_{11}^{80}$ | $1000001011_1^0$ | (1234610589711) | pA | 3/11 | [1/6,2/5] | 1/6 | 1/2 | 22 | 0.566 |
| $s_{11}^{81}$ | $1000001010_1^0$ | (1234697105811) | pA | 3/11 | [1/6,1/2] | 1/6 | 1/3 | 20 | 0.571 |
| $s_{11}^{82}$ | $1000001110_1^0$ | (1234698105711) | pA | 3/11 | [1/6,1/2] | 1/6 | 1/3 | 20 | 0.571 |
| $s_{11}^{83}$ | $1000001111_1^0$ | (1234589710611) | pA | 3/11 | [1/6,2/5] | 1/6 | 1/2 | 22 | 0.566 |
| $s_{11}^{84}$ | $1000001101_1^0$ | (1234581069711) | pA | 3/11 | [1/6,1/2] | 1/6 | 1/3 | 20 | 0.571 |
| $s_{11}^{85}$ | $1000001100_1^0$ | (1235791046811) | pA | 2/11 | [1/6,1/5] | 1/6 | 1/2 | 18 | 0.241 |
| $s_{11}^{86}$ | $1000000100_1^0$ | (1234681057911) | pA | 2/11 | [1/7,1/4] | 1/7 | 1/2 | 16 | 0.393 |
| $s_{11}^{87}$ | $1000000101_1^0$ | (1234571069811) | pA | 3/11 | [1/7,1/2] | 1/7 | 1/2 | 18 | 0.629 |
| $s_{11}^{88}$ | $1000000111_1^0$ | (1234569810711) | pA | 3/11 | [1/7,1/2] | 1/7 | 1/2 | 18 | 0.629 |
| $s_{11}^{89}$ | $1000000110_1^0$ | (1234579106811) | pA | 2/11 | [1/7,1/4] | 1/7 | 1/2 | 16 | 0.393 |
| $s_{11}^{90}$ | $1000000010_1^0$ | (1234568107911) | pA | 2/11 | [1/8,1/3] | 1/8 | 1/2 | 14 | 0.517 |
| $s_{11}^{91}$ | $1000000011_1^0$ | (1234567910811) | pA | 2/11 | [1/8,1/3] | 1/8 | 1/2 | 14 | 0.517 |
| $s_{11}^{92}$ | $1000000001_1^0$ | (1234567810911) | pA | 2/11 | [1/9,1/2] | 1/9 | 1/9 | 12 | 0.692 |
| $s_{11}^{93}$ | $1000000000_1^0$ | (1234567891011) | fo | 1/11 | [1/11] | 1/11 | 1/2 | 10 | 0 |